\documentclass[letterpaper,12pt]{article}
\RequirePackage{cmap} 
\RequirePackage[T2A]{fontenc}
\RequirePackage[utf8]{inputenc}
\RequirePackage[english]{babel}
\RequirePackage[hidelinks,pdfpagelabels=false]{hyperref}

\RequirePackage{cite}
\usepackage{comment}
\usepackage{physics}
\usepackage{amssymb}
\usepackage[ 
top=0.75in,
right=0.75in,
bottom=0.75in,
left=0.75in,
footskip=2em
]{geometry}
\usepackage{authblk}
\usepackage{fancyhdr}
\fancypagestyle{myfirstpage}{%
	\lhead{}
	\chead{}
	\rhead{INR-TH-2017-023}
	\lfoot{}
	\cfoot{}
	\rfoot{}
	
}

\usepackage{xifthen}

\def \ifempty#1{\def\temp{#1} \ifx\temp\empty }
\newcommand{\arxiv}[2]{arXiv:\href{https://arxiv.org/abs/#1}{#1\ifthenelse{\isempty{#2}}{}{~[#2]}}}
\newcommand{\doi}[1]{doi:\href{https://doi.org/#1}{#1}}
\newcommand{\email}[1]{\href{mailto:#1}{#1}}

\begin{document}
\title{No static sphericaly symmetric wormholes in Horndeski theory}

\author{O. A. Evseev\footnote{{\bf email:} \email{oa.evseev@physics.msu.ru}} }
\author{O. I. Melichev\footnote{{\bf email:} \email{olegmelichev@gmail.com}} }
\affil{Institute for Nuclear Research of the Russian Academy of Sciences, 60th October Anniversary Prospect 7a, Moscow, 117312, Russia}
\affil{Department of Particle Physics and Cosmology, Faculty of Physics, M.~V.~Lomonosov Moscow State University, Vorobyovy Gory, 1-2, Moscow, 119991, Russia}

\date{November 11, 2017}
\maketitle
\thispagestyle{myfirstpage}
\vspace{-5pt}

\begin{abstract}
We consider the Horndeski theory in four-dimensional space-time. We show that this theory does not admit stable, static, spherically symmetric, asymptotically flat, Lorentzian wormholes.
\end{abstract}

\section{Introduction}
The Horndeski theory (generalized Galileons plus gravity) is the most general theory with second derivatives in the Lagrangian leading to equations of motion of the second order. This theory is described by the following Lagrangian \cite{Horndeski:1974wa,Nicolis:2008in,Deffayet:2009wt,Deffayet:2011gz}:
\begin{equation}
\label{eq:lagrangian_horndeski}
\mathcal{L} = \sum_{i = 2}^{5} \mathcal{L}_i
\end{equation}
\begin{equation*}
  \begin{aligned}
    \mathcal{L}_2 &= K(\phi, X),\\
    \mathcal{L}_3 &= -G_3 (\phi, X) \square \phi \textrm{,} \\
    \mathcal{L}_4 &= G_4 (\phi, X) R + G_{4X} \left[ ( \square \phi )^2 - ( \nabla_\mu \nabla_\nu \phi )^2 \right]\\
    \mathcal{L}_5 &= G_5 (\phi, X) G_{\mu\nu}\nabla^\mu\nabla^\nu \phi - \frac{1}{6} G_{5X} \left[ (\square \phi)^3 - 3 \square \phi ( \nabla_\mu \nabla_\nu \phi )^2 + 2 ( \nabla_\mu \nabla_\nu \phi )^3 \right] \textrm{,}\\
  \end{aligned}
\end{equation*}
~\\
where $ X = - \frac{1}{2} \nabla_\mu \phi \nabla^\mu \phi$, $\Box \phi = \nabla_\mu \nabla^\mu \phi$, $R$ is the scalar curvature and $G_{\mu\nu} = R_{\mu\nu} - \frac{1}{2}g_{\mu\nu} R$ is the Einstein tensor, metric signature is $(-,+,+,+)$.

The interest to this theory stems, in particular, from the fact that it admits stable null energy condition (NEC) violating solutions \cite{Nicolis:2008in, Creminelli:2010ba, Goon:2011uw, Goon:2011qf, Kobayashi:2011nu, Rubakov:2014jja}. The latter property makes Galileons natural candidates for fields that may support Lorentzian wormholes \cite{Morris:1988cz, Morris:1988tu, wormholes_static, Hochberg:1998ha, Novikov:2007zz, Shatskiy:2008us} and/or semi-closed worlds \cite{Frolov:1988vj, Guendelman:2010pr, Chernov:2007cm, Dokuchaev:2012vc}. It has been shown, however, that both asymptotically flat, static, spherically symmetric wormholes \cite{Rubakov:2015gza, Rubakov:2016zah} and semiclosed worlds \cite{Evseev:2016ppw} are unstable in $\mathcal{L}_3$ theories with minimal coupling to gravity, {\it{i.e.}} $G_4 = {M_{Pl}^2}/{2}$, $G_5 = 0$.

The purpose of this paper is to extend this result to wormholes in the most general Horndeski theory~\eqref{eq:lagrangian_horndeski}. The proof of the analogous (by interchanging radial coordinate and time) no-go theorem for bouncing cosmologies was given in \cite{Libanov:2016kfc} and extended to the full Horndeski theory in \cite{Kobayashi:2016xpl}.

The paper is organized as follows. We give a brief review of some of the results obtained by T.~Kobayashi, H.~Motohashi and T.~Suyama \cite{Kobayashi:2012kh,Kobayashi:2014wsa} in Sec.~\ref{sec:stability}, as they are essential for our argument. We prove the instability in Sec.~\ref{sec:proof}.
\section{Stability conditions in terms of the Lagrangian functions}
\label{sec:stability}
	We consider static, spherically symmetric, asymptotically flat Lorentzian wormholes. With a convenient gauge choice, the general form of metric is
\begin{equation*}
ds^2 = - A(r)dt^2 + \frac{dr^2}{B(r)} + r^2 \left( d \theta^2 + \sin^2 \theta d \varphi^2 \right).
\end{equation*}
T.~Kobayashi, H.~Motohashi and T.~Suyama in \cite{Kobayashi:2012kh,Kobayashi:2014wsa} obtained the stability conditions for perturbations about this background by performing the analysis in terms of spherical harmonics within the Regge-Wheeler approach \cite{Regge:1957td,Zerilli:1970se,Motohashi:2011pw}.

The odd-parity sector, emerging due to the gravitational wave part of perturbations, gives three stability conditions:
\begin{align}
    \mathcal{F} &= 2\left(
G_4+\frac{1}{2}B\phi'X'G_{5X}-XG_{5\phi}
\right) > 0, \label{eq:def-f}\\
    \mathcal{G} &= 2\left[ G_4-2XG_{4X}
+X\left(\frac{A'}{2A}B\phi'G_{5X}+G_{5\phi}\right)
\right] > 0, \label{eq:def-g}\\
    \mathcal{H} &= 2\left[G_4-2X G_{4X}
+X\left(\frac{B \phi'}{r} G_{5X}+ G_{5\phi}\right)\right] > 0\text{,}\label{eq:def-h}
\end{align}
where prime denotes $d/dr$.
These are obtained by requiring the abscense of ghost and gradient instabilities for modes propagating along both radial and angular directions. In particular, the condition (\ref{eq:def-f}), which we use in what follows, is needed to avoid the gradient instability.

The even parity sector is of particular interest, as it involves both scalar field perturbations and gravity waves. It gives the following stability condition:
\begin{equation}
\label{eq:2p-f}
2 \mathcal{P}_1-\mathcal{F} > 0,
\end{equation}
where
\begin{align}
&\mathcal{P}_1 = \,\, \frac{B \left( 2r\mathcal{H}+\Xi\phi ' \right)}{2 A r^2\mathcal{H}^2} \cdot \frac{d}{dr} \left[ \frac{A r^4\mathcal{H}^4}{ \left( 2r\mathcal{H}+\Xi\phi ' \right)^2 B} \right],\\
&\Xi = \,\, 2r^2 \left[ - X G_{3X} + \frac{2 B \phi'}{r} \left\{ G_{4X} + 2 X G_{4XX}  - (X G_{5 \phi})_{X} \right\} \right.\nonumber\\
&\phantom{\Xi = \,\, 2r^2 \left[\right.}\left. + G_{4 \phi} + 2 X G_{4 \phi X} - \frac{1}{r^2} X G_{5X} + \frac{B}{r^2} (3 X G_{5X} + 2 X^2 G_{5XX} )\right].
\end{align}
The condition (\ref{eq:2p-f}) is necessary for the absence of ghosts.

\section{No-go theorem for Horndeski theory}
\label{sec:proof}

The stability conditions most relevant for our purposes are \eqref{eq:def-f} and \eqref{eq:2p-f}. We define a variable 
\begin{equation*}
Q=\sqrt{\frac{B}{A}}\,\frac{2r\mathcal{H}+\Xi\phi '}{r^2\mathcal{H}^2},
\end{equation*}
and write \eqref{eq:2p-f} in the following form:
\begin{equation*}
2\mathcal{P}_1-\mathcal{F}=-2\sqrt{\frac{B}{A}}\,\frac{Q'}{Q^2}-\mathcal{F}>0, 
\end{equation*}
or
\begin{equation}
\frac{Q'}{Q^2}<-\frac{1}{2}\sqrt{\frac{A}{B}}\,\mathcal{F}.
\end{equation}
By integrating this relation from $r$ to $r'>r$ we obtain (cf. \cite{Rubakov:2016zah})
\begin{equation}
\label{eq:cond-int}
Q^{-1}(r)-Q^{-1}(r')< - \int\limits_{r}^{r'}dr\sqrt{\frac{A}{B}}\mathcal{F}
\end{equation}
Now, let $Q^{-1}(r)$ be negative at some $r$. Then we write \eqref{eq:cond-int} as follows:
\begin{equation}
\label{eq:cond-minus-infty}
Q^{-1}(r')>Q^{-1}(r) + \int\limits_{r}^{r'}dr\sqrt{\frac{A}{B}}\mathcal{F},
\end{equation}
and notice that if the integral on the right side of the inequality \eqref{eq:cond-minus-infty} diverges as $r' \rightarrow + \infty$, then $Q^{-1}(r')$ has to become positive, meaning that $Q^{-1}(r^{*}) = 0$ at some point $r^{*}$ and $Q$ is singular at this point.

Conversely, let $Q^{-1}(r')$ be positive at some $r'$, then we write \eqref{eq:cond-int} as
\begin{equation}
\label{eq:cond-plus-infty}
Q^{-1}(r) < Q^{-1}(r') - \int\limits_{r}^{r'}dr\sqrt{\frac{A}{B}}\mathcal{F},
\end{equation}
and see that if the integral diverges as $r \rightarrow - \infty $ then $Q^{-1}(r)$ has to become negative, meaning again a singular $Q$ at some point $r^{*}$ where $Q^{-1}(r^{*}) = 0$.

For asymptotically flat wormhole, one has $A(r) \rightarrow 1$, $B(r) \rightarrow 1$ as $r \to \pm \infty$. Furthermore,  General Relativity is restored away from the wormhole throat provided that
\begin{equation}
\left\{\!\!
\begin{array}{l}
G_4 \rightarrow M_{Pl}^2/2 \\
G_5 \rightarrow 0 \\
\end{array}
\right. \text{at } r \rightarrow \pm \infty \text{.}
\end{equation}
Equation~\eqref{eq:def-f} then leads to $\mathcal{F}(r) \rightarrow M_{Pl}^2$ as $r \rightarrow \pm \infty$, so the integral in Eq.~\eqref{eq:cond-int} diverges as $r' \rightarrow + \infty$ and $r \rightarrow - \infty$. This completes the argument.

To make contact with Ref.~\cite{Rubakov:2016zah}, we notice that in the cubic Galileon theory with  $G_4 = M_{Pl}^2/2$, $G_5 = 0$, we have
\begin{equation*}
Q = \frac{\mathcal{Q}}{M_{Pl}^4}, \hspace{5mm} \mathcal{F} = M_{Pl}^2,
\end{equation*}
where $\mathcal{Q}$ is the variable introduced by V.~Rubakov in \cite{Rubakov:2016zah}. Thus, the inequality (\ref{eq:cond-int}) coincides with that used in \cite{Rubakov:2016zah}.

\section*{Acknowledgements}
The authors are indebted to V.~Rubakov, S.~Mironov and V.~Volkova for helpful discussions. This work has been supported by Russian Science Foundation grant 14-22-00161.

\end{document}